\documentclass[sigconf,screen]{acmart}
\acmConference[AIware 2024]{1st ACM International Conference on AIware}{July 2024}{Porto de Galinhas, Brazil}

\usepackage{enumitem}
\usepackage{ulem}
\usepackage{float}
\usepackage{soul}
\usepackage{xcolor}
\usepackage{titlesec}
\usepackage{subcaption} 

\usepackage[hyphens]{xurl} 
\usepackage[hyphens]{url}

\AtBeginDocument{%
  \providecommand\BibTeX{{%
    \normalfont B\kern-0.5em{\scshape i\kern-0.25em b}\kern-0.8em\TeX}}}

\begin{document}

\title{A Comparative Analysis of Large Language Models for Code Documentation Generation}





\author{Shubhang Shekhar Dvivedi}
\authornote{All authors contributed equally to this research.}
\email{shubhang20474@iiitd.ac.in}
\affiliation{%
  \institution{IIIT Delhi}
  \city{New Delhi}
  \country{India}
}

\author{Vyshnav Vijay}
\authornotemark[1]
\email{vyshnav20157@iiitd.ac.in}
\affiliation{%
  \institution{IIIT Delhi}
  \city{New Delhi}
  \country{India}
}

\author{Sai Leela Rahul Pujari}
\authornotemark[1]
\email{sai20401@iiitd.ac.in}
\affiliation{%
  \institution{IIIT Delhi}
  \city{New Delhi}
  \country{India}
}

\author{Shoumik Lodh}
\authornotemark[1]
\email{shoumik20407@iiitd.ac.in}
\affiliation{%
  \institution{IIIT Delhi}
  \city{New Delhi}
  \country{India}
}

\author{Dhruv Kumar}
\email{dhruv.kumar@iiitd.ac.in}
\affiliation{%
  \institution{IIIT Delhi}
  \city{New Delhi}
  \country{India}
}



\begin{abstract}
This paper presents a comprehensive comparative analysis of Large Language Models (LLMs) for generation of code documentation. Code documentation is an essential part of the software writing process. The paper evaluates models such as GPT-3.5, GPT-4, Bard, Llama2, and Starchat on various parameters like Accuracy, Completeness, Relevance, Understandability, Readability and Time Taken for different levels of code documentation. Our evaluation employs a checklist-based system to minimize subjectivity, providing a more objective assessment. We find that, barring Starchat, all LLMs consistently outperform the original documentation. Notably, closed-source models GPT-3.5, GPT-4, and Bard exhibit superior performance across various parameters compared to open-source/source-available LLMs, namely LLama 2 and StarChat. Considering the time taken for generation, GPT-4 demonstrated the longest duration, followed by Llama2, Bard, with ChatGPT and Starchat having comparable generation times. Additionally, file level documentation had a considerably worse performance across all parameters (except for time taken) as compared to inline and function level documentation.
\end{abstract}




\maketitle

\section{Introduction}
Code documentation refers to the process of describing the functionality, purpose, and usage of software code through accompanying text, comments, or annotations. High-quality code documentation is extremely important \cite{lethbridge_how_2003} as it serves as a vital communication tool for developers, aiding in the understanding, maintenance, and collaboration of software projects. Effective code documentation not only enhances code readability and comprehension but also allows software maintenance, debugging, and future development efforts. Code documentation plays an indispensable role in software engineering, promoting clarity, reliability, and efficiency in the development life-cycle.

In the field of artificial intelligence, Large Language Models (LLMs) are revolutionizing how we handle information and develop software \cite{noauthor_programmers_nodate}. These models, built on massive data sets and sophisticated algorithms, have moved beyond mere convenience to become vital in the software creation process, offering unprecedented assistance in code generation, debugging, and indeed, code documentation \cite{khan_automatic_2022}. As software grows more complex and development speeds up, our dependence on these digital assistants is increasing. This makes it crucial to comprehensively assess their effectiveness and accuracy.

Code documentation generation using LLMs \cite{rai_review_2022, ahmed_few-shot_2023} has the potential to speed up the software development process, as creating documentation is often considered as a tedious and time consuming process by most programmers \cite{shmerlin2015document}. LLMs can play a very crucial role in bridging this gap if they are able to create and maintain the code documentation in a consistent and efficient manner. To maintain the standards, the LLMs must be able to summarize the code, explain the code, and also be able to format the code into an easily understandable, legible and usable form.

Existing work on generating code documentation using LLMs has focused on studying a single LLM across multiple languages \cite{khan_automatic_2022}, comparing LLM generated documentation with student generated documentation and documentation outsourced from paid services \cite{noauthor_comparing_nodate, alizadeh_open-source_2023}. Yet, there is no existing study which provides a comprehensive comparison of various LLMs for generating code documentation across multiple paradigms like comparing across different documentations levels, comparing between open source and close source LLMs etc. This comparison is important because it can help the software developers to select the best LLM for the code documentation based on their requirements and constraints. This paper precisely bridges this research gap and presents a comprehensive comparitive analysis of LLMs for generation of code documentation. We evaluate models such as GPT-3.5 \cite{brown_language_2020}, GPT-4 \cite{openai_gpt-4_2023}, Bard \cite{manyika_overview_nodate}, Llama2 \cite{touvron_llama_2023}, and Starchat \cite{li_starcoder_2023} on various parameters like Accuracy, Completeness, Relevance, Understandability, Readability and Time Taken for different levels of code documentation (inline level, function level and file level). In order to minimize subjectivity, we use a checklist-based system providing a more objective assessment.

In terms of research questions, this paper aims to find answers to the following research questions:
\begin{itemize}
    \item \textbf{RQ1:} How do popular LLMs compare on their ability to generate various levels of code documentations?
    \item \textbf{RQ2:} How does the performance of LLMs compare with the human generated documentation for same codebases?
    \item \textbf{RQ3:} How does the the performance of closed/private sourced LLMs compare with LLMs whose sources are publicly available?
\end{itemize}

Our results indicate that all LLMs (except StarChat) consistently outperform the original documentation generated by humans. Our evaluation also reveals that closed-source models such as GPT-3.5, GPT-4, and Bard exhibit superior performance across various parameters compared to open-source/source-available LLMs, namely LLama 2 and StarChat. In terms of the time taken for generating the code documentation, GPT-4 is the slowest followed by Llama2, Bard, with ChatGPT and Starchat having comparable generation times. Finally, file level documentation had a considerably worse performance across all parameters (except for time taken) as compared to inline and function level documentation.

This paper aims to pave the path for future research and development, pinpointing areas where LLMs excel and identifying gaps that require attention in the realm of automating the task of code documentation using LLMs.
\section{Related Work}
Code documentation generation has been a subject of considerable research in the previous years, especially since the rise of LLMs \cite{khan_automatic_2022, noauthor_comparing_nodate, alizadeh_open-source_2023} and the rapid improvement in code comprehensibility softwares using AI based techniques \cite{su_hotgpt_2023, macneil_generating_2022}.

Existing approaches to automatic generation of code documentation use data publishing frameworks \cite{che_automatic_nodate} or transformer-based approaches \cite{meneses_documentation_nodate}. A recent study done by Zhu et al \cite{10.1145/3631975} compares the ability of creating code summaries using Information Retrieval methods and Deep Learning methods (including one open-source LLM, i.e. StarCoder \cite{li_starcoder_2023}). 


Several studies have been done on the ability of LLMs to perform tasks related to code documentation \cite{su_hotgpt_2023, macneil_generating_2022, geng2023large, noauthor_programmers_nodate}. Su et al \cite{su_hotgpt_2023} propose using LLMs for key information extraction from the source code.  Macneil et al \cite{macneil_generating_2022} propose creating code summaries from code in the context of CS education. Geng et al \cite{geng2023large} study multi-intent comment generation capabilities (like code functionality, how to use the code etc.) of LLMs. There have also been attempts of studying conversational uses of LLMs as a coding assistant, being able to translate, explain, and complete code as it is being written \cite{noauthor_programmers_nodate}.

With regards to code documentation generation using LLMs, Khan et al \cite{khan_automatic_2022} studied the performance of GPT-3 across 6 programming languages. They found that Codex (a GPT-3 based model pretrained on both natural and programming language) outperformed the state of the art documentation generation techniques present till 2022. Empirical studies have also been conducted to compare LLM generated code documentation with student generated documentation \cite{noauthor_comparing_nodate} which found that LLM generated documentations were significantly easier to understand and more accurate. Alizadeh et al compared the performance of ChatGPT (GPT-3.5), with some open source LLMs (like HugginChat and FLAN) and documentation generated by human based paid services (MTurk) \cite{alizadeh_open-source_2023}. They found that the LLMs outperform MTurk and the open source LLMs demonstrate competitive potential against ChatGPT in certain tasks.



All of the above studies have consistently demonstrated promising outcomes in the realm of automatic code documentation and associated tasks. 

However, in comparison to the above mentioned existing research, our study offers new insights from multiple perspectives: (1) We present a comprehensive analysis of multiple LLMs (both closed-source and open-source LLMs).
(2) Our study analyzes multiple code documentation levels (inline level, function level, class level). 
Thus, our study further enriches the understanding of LLM capabilities in code documentation.

\section{Methodology}
\subsection{Levels of documentation}

To conduct a thorough analysis of the LLMs' capabilities in generating code documentation, we decided to categorise the documentations into four distinct categories/levels \cite{rai_review_2022}:
\begin{itemize}[leftmargin=*]
  \item Inline level documentation
  \item Function level documentation
  \item Class level documentation
  \item Folder level documentation
\end{itemize}

Each level represents a different scope and complexity of documentation, providing insights into how well each LLM handles varying degrees of detail and abstraction in technical writing.

\subsubsection{\textbf{Inline Level:}}

It involves the generation of comments within the code, explaining specific lines or blocks of code. Inline documentation is crucial for clarifying complex or non-obvious parts of the code, enhancing readability and maintainability. It can also be very helpful in explaining a piece of code line by line to new coders. This will help evaluate the LLMs' ability to produce concise, relevant, and helpful inline comments that enhance understanding without cluttering the code.

\subsubsection{\textbf{Function level:}}

At the function level, the focus shifts to documenting individual functions or methods. Good function-level documentation explains the purpose, parameters, return values, exceptions, and usage examples of functions. This assessment will measure how well each LLM can summarise the functionality and provide clear, accurate descriptions for function usage.

\subsubsection{\textbf{Class level:}}

Class level documentation provides an overview of a class, its purpose, and its interactions with other classes. It often includes descriptions of key attributes and methods within the class. This level of documentation is critical for understanding the design and architecture of the software. 

\subsubsection{\textbf{Folder Level:}}

Folder Level Documentation is the highest level of documentation involving creating overviews for entire folders or modules, summarising the functionalities and interactions of multiple classes or files within them. We did not perform the analysis on folder level documentation because of technological limitations in most of the LLMs (except for GPT-4) which could not handle multiple file inputs at the same time.

\subsection{Dataset} \label{sec:dataset}

A set of 14 python code snippets were selected from a list of well documented publicly available repositories \cite{githubGitHubEmperorRPDataEvaluationsComparativeAnalysisofLLMsforCodeDocumentationGeneration}. These selections were made to ensure a comprehensive analysis across different levels of documentation, including inline, function, and file levels.

For inline level documentation, our analysis included the following code repositories:

1 - verify.py from Bitcoin \cite{githubBitcoincontribverifybinariesverifypyE25af11225d9d94ecf7068bf7a9a359268786fbe},

2 - clip$\_$ops.py from TensorFlow \cite{githubTensorflowtensorflowpythonopsclip_opspyV2130}, and

3 - conftest.py \cite{githubGitHubSqlalchemysqlalchemy}, test\_functions.py \cite{githubSqlalchemytestsqltest_functionspyMain}, and basic\_association.py from SQLAlchemy \cite{githubSqlalchemyexamplesassociationbasic_associationpy8503dc2e948908199cd8ba4e6b1d1ddcf92f4020}.

For function level documentation, we focused on functions like:

1 - update\_prediction\_data() from the Reddit archive \cite{githubRedditr2r2libinventorypy753b17407e9a9dca09558526805922de24133d53}, 

2 - test\_flush\_no\_pk() from SQLAlchemy \cite{githubSqlalchemyexamplesperformancebulk_insertspy8503dc2e948908199cd8ba4e6b1d1ddcf92f4020},

3 - fit() from Scikit-Learn \cite{githubScikitlearnsklearncluster_kmeanspy92c9b1866fab77412d8fe93cb3716c03d80ad8ed}, and

4 - check\_multisig() from Bitcoin \cite{githubBitcoincontribverifybinariesverifypyE25af11225d9d94ecf7068bf7a9a359268786fbe}.

File level documentation analysis utilized files from:

1 - bench\_glmnet.py from Scikit-Learn \cite{githubScikitlearnbenchmarksbench_glmnetpy1495f69242646d239d89a5713982946b8ffcf9d9},

2 - compat.py from TensorFlow \cite{githubTensorflowtensorflowpythonutilcompatpyV2130},

3 - message-capture-parser.py from Bitcoin \cite{githubBitcoincontribmessagecapturemessagecaptureparserpyE25af11225d9d94ecf7068bf7a9a359268786fbe}, and

4 - dict\_of\_sets\_with\_default.py \cite{githubSqlalchemyexamplesassociationdict_of_sets_with_defaultpy8503dc2e948908199cd8ba4e6b1d1ddcf92f4020}, and basic\_association.py from SQLAlchemy \cite{githubSqlalchemyexamplesassociationbasic_associationpy8503dc2e948908199cd8ba4e6b1d1ddcf92f4020}.

The selection of codebases for analysis was based on their popularity and the perceived quality of documentation within the software development community. We chose codebases known for their widespread use and reputation for excellent documentation, as indicated by discussions and recommendations in online forums, technical communities, and software development platforms. Focusing on codebases with well-regarded documentation allows for better assessment of code documentation generation methods. This approach ensures credibility and relevance in the evaluation process, strengthening the validity and usefulness of the research findings.

These code snippets were then individually run through all the LLMs. There were a total of 84 documentations (14 code snippets * 6 documentations each = 84 documentations) and each of them were evaluated on all the metrics (namely accuracy, completeness, relevance, understandability, readability and time taken) and the evaluated data was then collected for further analysis \cite{githubGitHubEmperorRPDataEvaluationsComparativeAnalysisofLLMsforCodeDocumentationGeneration}.

\subsection{Evaluation parameters and metrics} \label{sec:metrics}

We carefully picked evaluation criteria and their measures to thoroughly analyse the database. Our goal was to choose the metrics in such a manner such that the evaluation of the documentations would be thorough and cover all the major aspects of a documentation. Importantly, we wanted our assessment to be as objective as possible, so the metrics we chose are designed to give a fair and straightforward evaluation of the documentation. In order to make the evaluations more objective and reduce biases, we created a checklists for some metrics (namely completeness, relevance and readability) and the evaluation on those metrics is done according to the checklists. The evaluation parameters and their corresponding metrics are listed down below:
\begin{enumerate}[leftmargin=*]
\item{\textbf{Accuracy:}}

Accuracy is the measure of how correctly the documentation describes the code. This metric checks if the statements generated in the documentation is factually true or not, given the source code as the context.
The accuracy is rated on a scale of 1-3, with 1 being the least accurate and 3 being perfectly accurate.

\item{\textbf{Completeness:}}

Completeness of the documentation indicates the extent to which all the (important) parts of the code have been covered or not.
Completeness is rated using a 5 point checklist on a scale of 0-5, on the basis of the number of checkpoints the documentation satisfied. Following are the checklist items for the various documentation levels:

\begin{itemize}[leftmargin=*]
\item{\textbf{For file level:}}
\begin{itemize}
    \item Brief description of the file/script’s purpose is included
    \item List of dependencies is provided
    \item Instructions or notes on how to use the script are included
    \item List of main functions or classes defined in the file (if applicable)
    \item Any special notes or considerations relevant to the script (e.g., context of use, limitations)
\end{itemize}

\item{\textbf{For function level:}}
\begin{itemize}
    \item The description of the function’s purpose is clear
    \item All parameters are described with type and purpose
    \item Return values are explained (if any)
    \item Exceptions or errors the function may raise are documented.(if any)
    \item Any additional notes or warnings are included
\end{itemize}

\item{\textbf{For inline level:}}
\begin{itemize}
    \item Complex code blocks have accompanying comments
    \item Non-obvious algorithmic decisions are explained (if any)
    \item Workarounds or technical debt is noted (if any)
    \item Any assumptions in the code are stated
    \item Deprecated methods or upcoming changes are flagged. (depends on LLM)
\end{itemize}
\end{itemize}
\item{\textbf{Relevance:}}
Relevance shows how relevant the generated documentation is to the actual code, essentially it's ability to stick to the subject matter of the code and not go off topic. Relevance is different from accuracy in terms of focusing on the alignment of documentation with the subject matter of the code.
Relevance is rated on a scale of \textbf{1} to \textbf{4} as explained below:

\textbf{1} - No relevant information; completely off-topic or non-existent.

\textbf{2} - Somewhat relevant; mixes essential and non-essential information.

\textbf{3} - Mostly relevant; includes most of the key details necessary for understanding the function.

\textbf{4} - Fully relevant; every piece of information helps understanding the function.

\item{\textbf{Understandability:}}
Understandability indicates how well a person reading the documentation is able to understand what the piece of code means/ how to use the code. Rating understandability is very much dependent on the level of experience and skill a user has in software development. Hence, there is a scope of understandability being a little subjective as an evaluation parameter, and is evaluated on the scale of \textbf{1} to \textbf{4}, where each value means the following:

\textbf{1} - Completely unintelligible; impossible to understand.

\textbf{2} - Somewhat understandable; could benefit from clearer language or examples.

\textbf{3} - Mostly understandable; fairly easy to grasp but could be improved.

\textbf{4} - Fully understandable; extremely clear and easy to understand, even for someone unfamiliar with the topic.

\item{\textbf{Readability:}}
Readability rates the formatting of the documentation. A well formatted documentation makes it convenient for the reader to go through the documentation.
Readability is rated using a \textbf{5}-point checklist on a scale of \textbf{0-5}, on the basis of the number of checkpoints the documentation satisfies. The checkpoints are given below:

\begin{itemize}[leftmargin=*]
\item{\textbf{File level:}}
Header comment has clear delineation (e.g., lines or asterisks). (Documentation is clearly separated from the code),
Consistent indentation is used for nested information,
Lines are properly aligned (horizontal scrolling is absent),
Is metadata like author, date, etc. present,
Presence of spacing between Sections.

\item{\textbf{Function level:}}
Is the docstring clearly distinguishable from the code,
Is there consistent formatting in the docstring (e.g., for parameters, returns),
Any examples within the docstring are properly indented,
Line breaks are used to separate different sections within the docstring,
Is the docstring structured logically (e.g., description, parameters, returns, examples)?

\item{\textbf{Inline level:}}
Are comments placed close to the code they describe or do comments follow the same indentation as the code they are describing,
Comments are brief and do not exceed the length of the code line,
Consistent style is used for single-line and multi-line comments,
Is there adequate spacing around comments for clarity,
Are comments placed logically in the code for easy association with the relevant code?
\end{itemize}

\item{\textbf{Time Taken:}}
This is a simple measure of the time taken by the model to fully generate the response after the prompt has been entered.
\end{enumerate}

\subsection{Prompts} \label{sec:prompts}

The prompts used as inputs to the LLMs were carefully constructed using the concepts of prompt engineering to generate the best possible responses from the large language models. The same prompt was used across models for the same level of documentation to maintain homogeneity and fairness while comparing the performances of models. The prompts were generally structured in a way such that they included the following:
\begin{itemize}[leftmargin=*]
\item{\textbf{Persona:}}
The LLM was assigned a persona
\item{\textbf{Context:}}
A context was given to the LLM under which the documentation is to be generated
\item{\textbf{Task:}}
The LLM was given the task using clear action words.
\item{\textbf{Format:}}
The LLM was specified the format in which the documentation was to be generated. 
\end{itemize}

The prompts used for each level of documentation were slightly different, and are given below:

\noindent \textbf{File level:}
\textit{As a code documentation assistant, you are responsible for documenting at the file/script level of the given code snippet. When provided a file level code, your approach involves adding a header comment at the top of the file. This comment should be the documentation for the code and include all relevant information needed to understand or use the script. Code is as follows: <insert code>}

\noindent \textbf{Function level:}
\textit{As a code documentation assistant, you are programmed to document at the function level of the given code snippet. Your approach involves placing comments directly under the def statement of the function. The output should be the entire code along with the documentation of the function written. Code is as follows: <insert code>}

\noindent \textbf{Inline level:}
\textit{As a code documentation assistant, you are assigned to document at the in-line level of the given code snippet. When in-line comments are needed, you insert comments within the code itself. The output should be the entire code, along with the documentation you’ve added. Code is as follows: <insert code>}

\subsection{LLMs Used}

In our comparative analysis of code documentation generation capabilities among various large language models (LLMs), both open-source/source-available and closed-source models were utilized. The LLMs included in our study are GPT-3.5, GPT-4, Bard, LLama2, and Starchat.

The closed source proprietary LLMs used are the following:
\begin{itemize}
\item{\textbf{GPT-3.5 \cite{brown_language_2020}:}}
Developed by OpenAI, GPT-3.5 is an advanced iteration of the Generative Pre-trained Transformer models, boasting around 175 billion parameters. It is one of the most popular LLMs used as of right now.

\item{\textbf{GPT-4 \cite{openai_gpt-4_2023}:}}
Also from OpenAI, GPT-4 is a more sophisticated and larger model than its predecessor, with improved performance in nuanced language understanding and generation. It is built on 1.76 trillion parameters and is also able to handle visual inputs along with textual ones.

\item{\textbf{Bard \cite{manyika_overview_nodate}:}}
Bard is a closed-source LLM developed by Google and is built on 137 billion parameters. It is one of the major players in the LLM space after OpenAIs chatGPT.
\end{itemize}

The open source/source available LLMs used are the following:
\begin{itemize}
\item{\textbf{LLama2 \cite{touvron_llama_2023}:}}
LLama2 is a source-available LLM developed by Meta. It has different model sizes with 7, 13 and 70 billion parameters. The one used for this study has 70 billion parameters.

\item{\textbf{Starchat \cite{li_starcoder_2023}:}}
Starchat 2 is a fine tuned version of the StarCoder code completion focused model and is one of the more commonly used code focused open source LLM. It is built on around 15 billion parameters.
\end{itemize}

\subsection{Data Analysis}

A total of 14 code snippets (including snippets of inline level, function level and file level) were collected from various code bases as mentioned in subsection \ref{sec:dataset}. These code snippets were then used to generate documentation from each of the 5 LLMs (GPT-3.5, GPT-4, Bard, LLama 2, StarChat) using the prompts corresponding to the documentation level as mention in subsection \ref{sec:prompts}.

All the generated documentations (14 documentations * 5 LLMs = 70 documentations) along with the original documentations of the 14 code snippets (14 documentations) were then evaluated by 2 of the authors of the paper. The evaluators have a good knowledge about the practice of source code documentations and have themselves created several documentations for various projects.

To avoid potential effects of the evaluator's biases on the evaluation of the datasets, percentage of disagreement was used for checking inter-rater reliability \cite{GISEV2013330}. Both the evaluators rated 6 code snippets, i.e. 6 snippets * 6 documentations each = 36 documentations, on 5 metrics each, namely accuracy, completeness, relevance, understandability and readability (time taken was not used as it was a temporal measure and can not be effected by biases). This makes it a total of 180 evaluations (36 documentations * 5 metrics = 180 evaluations). Out of these 180 evaluations, there were disagreements on only 21 of them, which makes the disagreement score to be 11.67 percent which is a fairly low disagreement score. Disagreement score is calculated as disagreement score = (disagreements/total evaluation)*100.

The low disagreement score is partly due to the checklist system used to evaluate the documentations on different metrics. It is also because of the uniformly high quality documentations generated by most of the LLMs most of the time. Most disagreements (13 out of 21) were present in the "understandability" metric due to the subjectivity of the metric.

The evaluators rated the each documentation on the 6 mentioned metrics (namely accuracy, completeness, relevance, understandability, readability and time taken) in accordance with the checklist system and rating systems as described in section \ref{sec:metrics} \cite{githubGitHubEmperorRPDataEvaluationsComparativeAnalysisofLLMsforCodeDocumentationGeneration}.

The evaluated data was then visualized and analysed python libraries like pandas, numpy, matplotlib and seaborn. The code for the data analysis can be accessed here \cite{githubGitHubEmperorRPDataEvaluationsComparativeAnalysisofLLMsforCodeDocumentationGeneration}. The findings and the visualizations are reported in the section \ref{sec:evaluation}.
\section{Evaluation} \label{sec:evaluation}
We conducted a thorough analysis of 14 Python code snippets (as mentioned in Section \ref{sec:dataset}), each accompanied with its respective documentation (generated by humans) for reference and comparison. We then ran various forms of analyses over the dataset, following are the results and observations:

\subsection{Overall Analysis}

We compute each metric (Accuracy, Completeness, Relevance, Understandability, Readability and Time Taken) for all the datasets and generated an output from each LLM. We then compute the metric by considering all the pairs of dataset and LLM we used for evaluation. For example, for computing the average accuracy in Table \ref{tab:stats}, we compute the average of accuracy achieved across all the pairs of dataset and LLM.

Table 1 presents a summary of the descriptive statistics like mean, standard deviation, range and distribution of values etc. for various parameters, including Accuracy, Completeness, Relevance, Understandability, Readability, and Time Taken.

\begin{table*}[h]
\centering
\begin{tabular}{|l|r|r|r|r|r|r|}
\hline
\textbf{Metric} & \textbf{Accuracy} & \textbf{Completeness} & \textbf{Relevance} & \textbf{Understandability} & \textbf{Readability} & \textbf{Time Taken} \\ \hline
mean               & 2.860          & 4.192              & 3.833           & 3.551                   & 4.525             & 16.977                \\ \hline
std                & 0.445          & 1.206              & 0.567           & 0.766                   & 0.935             & 17.994                \\ \hline
min                & 0.000          & 0.000              & 0.000           & 0.000                   & 0.000             & 0.000                 \\ \hline
25\%               & 3.000          & 4.000              & 4.000           & 3.000                   & 4.000             & 0.000                 \\ \hline
50\%               & 3.000          & 5.000              & 4.000           & 4.000                   & 5.000             & 11.000                \\ \hline
75\%               & 3.000          & 5.000              & 4.000           & 4.000                   & 5.000             & 30.250                \\ \hline
max                & 3.000          & 5.000              & 4.000           & 4.000                   & 5.000             & 75.000                \\ \hline
\end{tabular}
\caption{Accuracy, Completeness, Relevance, Understandability, Readability and Time Taken averaged across all LLMs}
\label{tab:stats}
\end{table*}

\subsection{Comparison across different LLMs}

To enhance the clarity of our findings, we utilized bar graph visualizations to compare mean ratings across different language models for each parameter.
\begin{figure}
  \centering
  \includegraphics[width=0.4\textwidth]{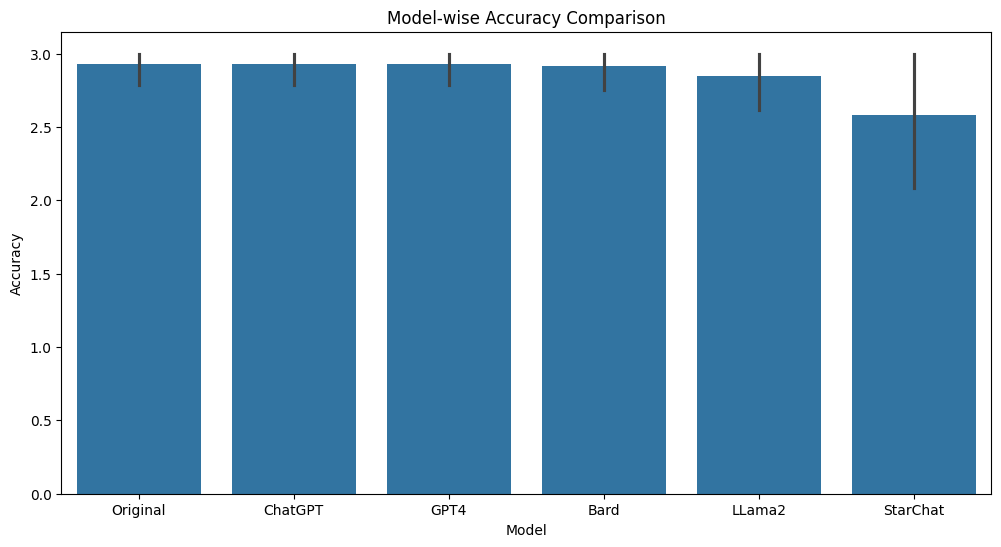}
  \caption{Model-wise accuracy comparison}
  \label{fig:model-accuracy}
\end{figure}

In figure 1, we can see that there is a significant drop in average accuracy in StarChat, meanwhile the rest of the LLMs have almost on par accuracy as compared to the original documentation.

\begin{figure}
  \centering
  \includegraphics[width=0.4\textwidth]{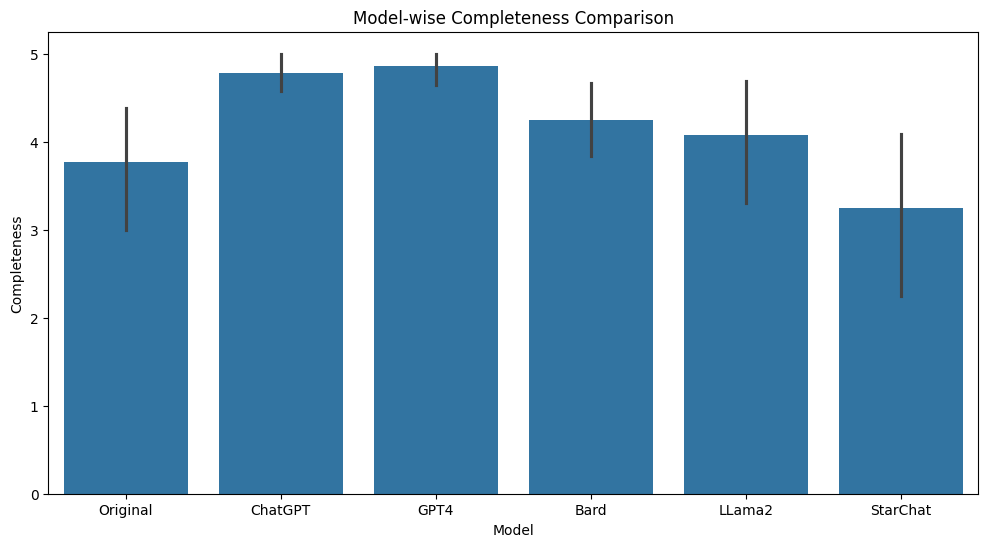}
  \caption{Model-wise completeness comparison}
  \label{fig:model-completeness}
\end{figure}

In figure 2, we can see that GPT-3.5 and GPT significantly outperform even the original documentation in terms of completeness and thoroughness of the code documentation, meanwhile Bard and LLama 2 have somewhat the same, and slightly better and StarChar has slightly worse performance in terms of completeness than the original documentation.

\begin{figure}
  \centering
  \includegraphics[width=0.4\textwidth]{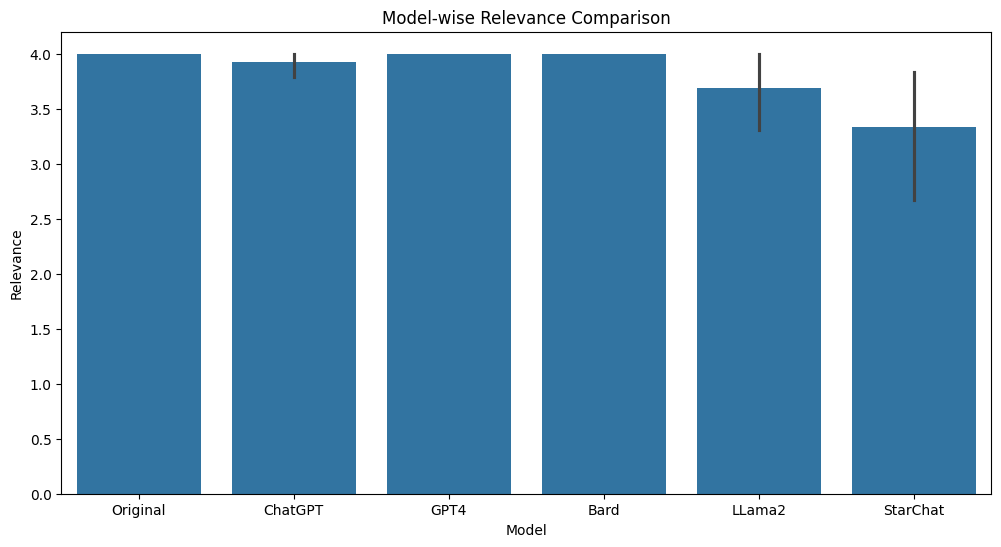}
  \caption{Model-wise relevance comparison}
  \label{fig:model-relevance}
\end{figure}

In figure 3, LLama 2 and StarChat have a slightly worse performance as compared to GPT-3.5 and 4 and Bard, which are on par with the original documentation.

\begin{figure}
  \centering
  \includegraphics[width=0.4\textwidth]{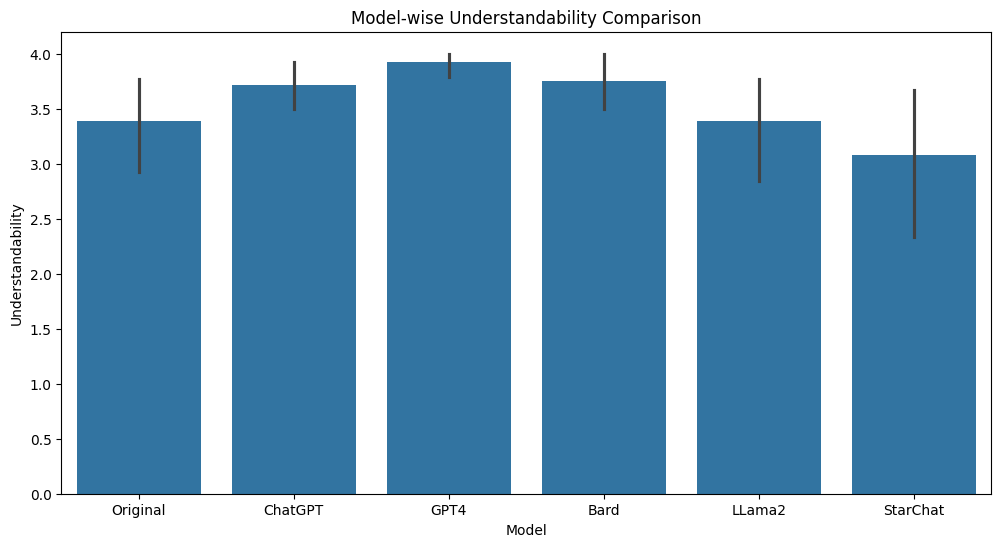}
  \caption{Model-wise understandability comparison}
  \label{fig:model-understandability}
\end{figure}

From figure 4, we can see that the documentation generated by GPT-3, GPT-4 and Bard have better understandability as compared to the original documentation. LLama 2 and StarChat have a somewhat similar performance as the original documentation.

\begin{figure}
  \centering
  \includegraphics[width=0.4\textwidth]{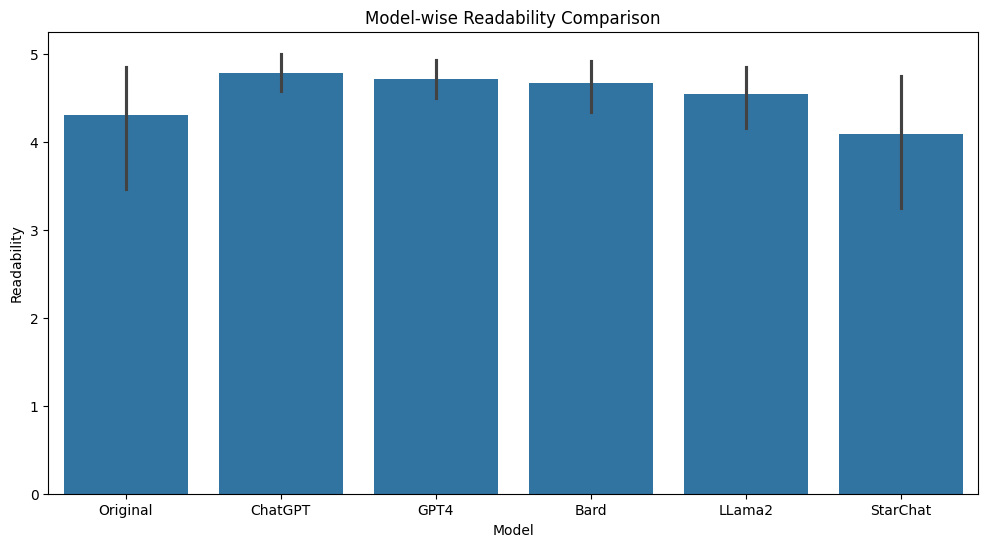}
  \caption{Model-wise readability comparison}
  \label{fig:model-readability}
\end{figure}

From figure 5, we can see that all the models have high Readability, and most of them outperform the original documentation by small margins.

\begin{figure}
  \centering
  \includegraphics[width=0.4\textwidth]{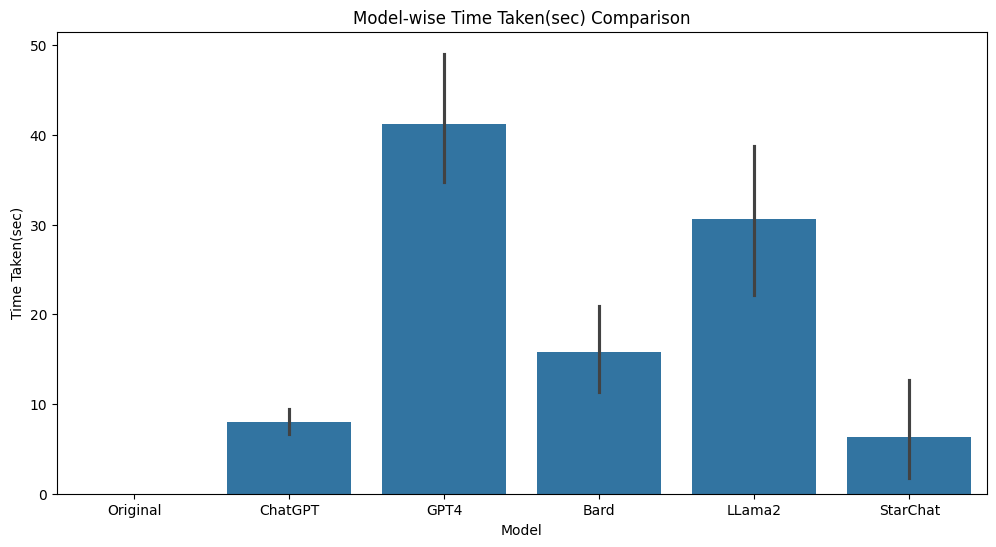}
  \caption{Model-wise Time taken comparison}
  \label{fig:model-time}
\end{figure}

From figure 6, we can say that the time taken by GPT-4 to generate the documentation is very high as compared to the other models. GPT-3.5 and StarChat generate the documentation in the shortest amount of time. Bard and LLama 2 lie somewhere in between. This is the only criteria observed where GPT-4 has performed the worst by a large margin.

\subsection{Correlation between different metrics}

\begin{figure}
  \centering
  \includegraphics[width=0.45\textwidth]{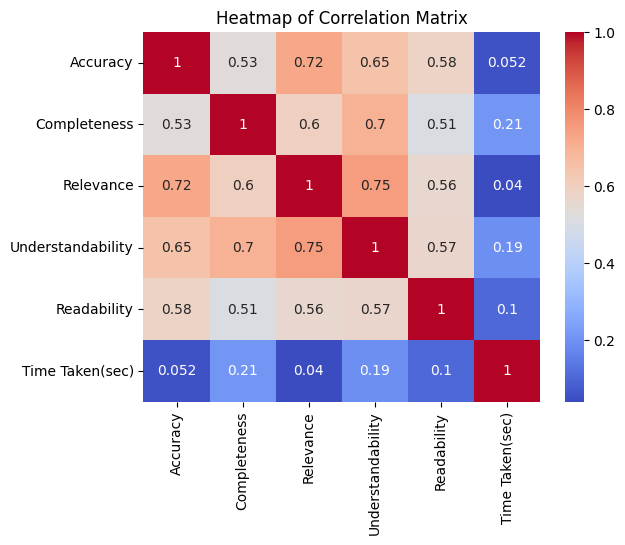}
  \caption{Metrics correlation heatmap}
  \label{fig:correlation}
\end{figure}

As we can see from the correlation heatmap in figure 7, there is a somewhat strong correlation between Accuracy, Understandability and Relevance (each about 0.7-0.75). Other than this, there is not strong correlation between the other parameters. It is also interesting to note that the time taken by each LLM is completely independent of all the other parameters.

\begin{figure}
  \centering
  \includegraphics[width=0.4\textwidth]{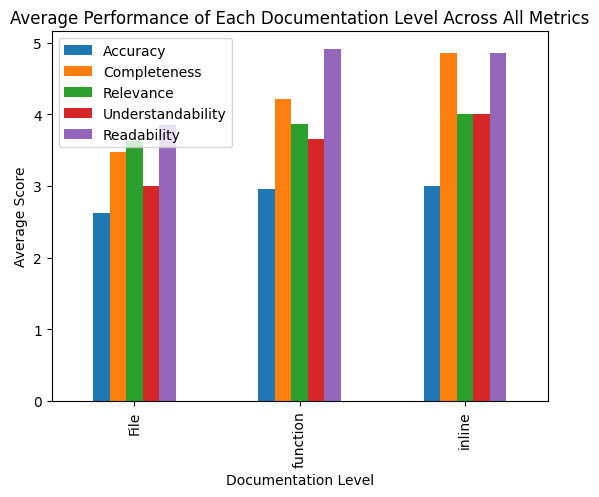}
  \caption{Documentation level-wise metrics comparison}
  \label{fig:level-parameters}
\end{figure}

\begin{figure}
  \centering
  \includegraphics[width=0.4\textwidth]{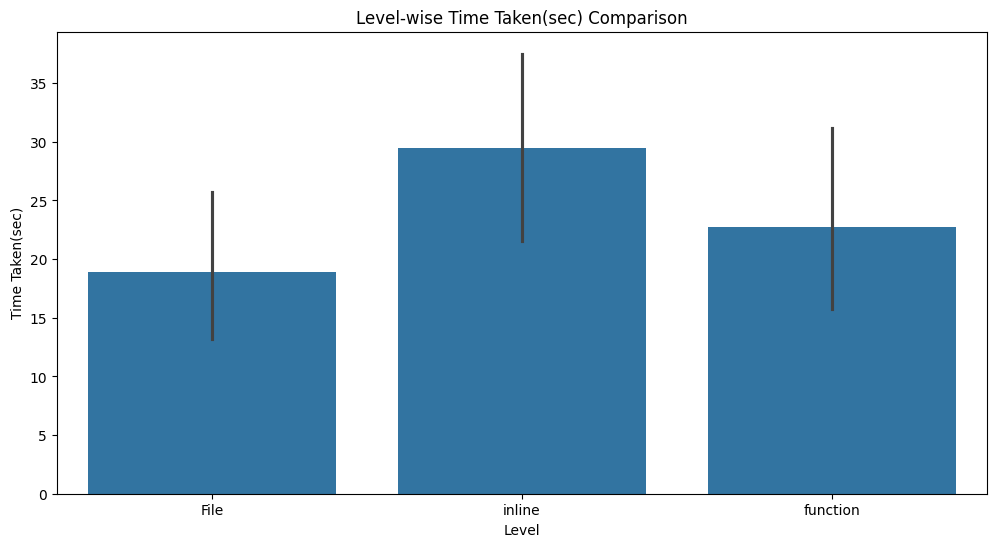}
  \caption{Documentation level-wise time taken comparison}
  \label{fig:level-time}
\end{figure}

\subsection{Comparison across different documentation levels}

After looking at and comparing the performances of each LLM across different metrics, we will now look at the general performance of all the LLMs across different documentation levels, namely inline level documentation, function level documentation and file level documentation.

From figure 8, we can see that performance of the LLMs while generating file level documentation has a consistently worse performance as compared to the performance while generating inline and function level documentation in all parameters.

From figure 9, it can be seen that even though file level documentation generated is rated less, it still takes the least amount of time and documentation is generated faster than the other 2 levels of documentation.

\subsection{Statistical Analysis}

For statistical analysis, ANOVA (Analysis of Variance) \cite{sthle_analysis_1989} is employed to assess the significance of differences among group means in multi-group comparisons. This method is particularly essential when evaluating multiple models across various parameters, such as performance metrics. ANOVA utilizes F-statistics and p-values to determine whether observed variances across models are statistically significant or occur by chance, thereby providing a rigorous and technical approach to model evaluation and comparison. The alpha level or significance level used for this analysis is 0.05.

\begin{table}[h]
\centering
\begin{tabular}{|l|r|r|}
\hline
\textbf{Criterion} & \textbf{F-statistic} & \textbf{P-value} \\ \hline
Accuracy           & 1.1803               & 0.3271           \\ \hline
Completeness       & 3.9874               & 0.0030           \\ \hline
Relevance          & 3.1599               & 0.0123          \\ \hline
Understandability  & 2.2788               & 0.05562          \\ \hline
Readability        & 1.0665               & 0.3861           \\ \hline
Time Taken         & 29.5365              & $1.0497 \times 10^{-16}$ \\ \hline
\end{tabular}
\caption{Statistical Analysis Results}
\label{tab:stats_results}
\end{table}

\begin{itemize}[leftmargin=*]
\item {\textbf{Accuracy:}}

The F-statistic is relatively low, and the p-value is higher than the common alpha level of 0.05. This suggests that there is not enough statistical evidence to conclude that the model significantly affects accuracy.

\item {\textbf{Completeness:}}

The F-statistic is higher, and the p-value is below 0.05. This indicates a statistically significant effect of the model on completeness.

\item {\textbf{Relevance:}}

A higher F-statistic and a p-value below 0.05 suggest a significant effect of the model on relevance.

\item {\textbf{Understandability:}}

The F-statistic indicates some effect, but the p-value is slightly above 0.05. This suggests that the model's impact on understandability is borderline significant; it's close to being significant but doesn't quite reach the conventional threshold.

\item {\textbf{Readability:}}

Both the F-statistic and the p-value indicate that the model does not have a significant effect on readability.

\item {\textbf{Time Taken:}}

The very high F-statistic combined with an extremely low p-value (practically zero) strongly suggests that the model has a significant effect on the time taken.
\end{itemize}

\section{Discussion}
Our comparative analysis of code documentation generated by various Large Language Models (LLMs) yielded several significant findings. Firstly, with the exception of Starchat, all LLMs demonstrated either equivalent or superior performance compared to the original documentation, highlighting their potential for automating documentation tasks. Secondly, closed-source models like GPT-3.5, GPT-4, and Bard consistently outperformed their open-source counterparts, Llama2 and Starchat, across a majority of evaluation parameters. Notably, GPT-4 exhibited the longest time taken for generation, followed by Llama2 and Bard, while ChatGPT and Starchat had comparable times. Lastly, file-level documentation consistently displayed poorer performance across all parameters (except for time taken) when compared to inline and function-level documentation.

Following are some implications, recommendations, and suggestions for the software development process in accordance with the findings of our paper:

\begin{enumerate}[leftmargin=*]
\item \textbf{Performance Discrepancies between LLMs:}
The varying performance of LLMs underscores the importance of carefully selecting the model based on specific project requirements and evaluation criteria. Software developers should consider conducting pilot tests with multiple LLMs to determine which model best suits their documentation needs.

\item \textbf{Closed-Source vs. Open-Source Models:}
The superiority of closed-source models suggests potential benefits in investing in proprietary LLMs for documentation tasks.

\item \textbf{Time Taken for generation considerations:}
The variation in generation time among LLMs highlights the importance of balancing performance with efficiency in documentation tasks. Developers should weigh the trade-offs between documentation quality and generation time when selecting LLMs for their projects. Research efforts should focus on developing strategies to optimize generation time without compromising documentation quality, enhancing the overall efficiency of LLM based documentation workflows.

\item \textbf{File-Level Documentation Challenges:}
The comparatively inferior performance of file-level documentation emphasizes the need for alternative approaches to improve its effectiveness. Future research could explore innovative methods for enhancing the quality and utility of file-level documentation, such as incorporating contextual information or utilizing specialized LLM architectures fine-tuned for larger code structures.
\end{enumerate}
\section{Conclusion}
Based on our comparative analysis, some very interesting findings have emerged. Except for Starchat, all Large Language Models (LLMs) demonstrate either equivalent or superior performance when compared to the original documentation, with Starchat consistently yielding suboptimal results. Secondly, closed-source models such as GPT-3.5, GPT-4, and Bard consistently outperform their open-source counterparts, Llama2 and Starchat, across a majority of evaluation parameters. Notably, in terms of generation time, GPT-4 exhibits the longest duration, followed by Llama2 and Bard, while ChatGPT and Starchat have comparable generation times. Lastly, file-level documentation displays significantly poorer performance across all parameters (except for time taken) as compared to inline and function-level documentation.
\bibliographystyle{ACM-Reference-Format}
\bibliography{references.bib}

\appendix

\end{document}